\documentclass[lettersize,journal, twocolumn]{IEEEtran}
\usepackage{amsmath,amsfonts}
\usepackage{algorithmic}
\usepackage{algorithm}
\usepackage{array}
\usepackage[caption=false,font=normalsize,labelfont=sf,textfont=sf]{subfig}
\usepackage{textcomp}
\usepackage{stfloats}
\usepackage{url}
\usepackage{verbatim}
\usepackage{graphicx}
\usepackage{cite}
\usepackage{comment}
\usepackage{booktabs} 
\usepackage[hidelinks]{hyperref}

\title{Internet of Intelligent Reflecting Surfaces (IoIRS)}

\newcommand{\orcidiconFeb}{\href{https://orcid.org/0009-0008-2632-1140}{\includegraphics[scale=0.1]{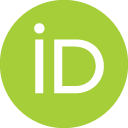}}}
\newcommand{\orcidiconAso}{\href{https://orcid.org/0009-0002-8206-0463}{\includegraphics[scale=0.1]{orcidID128.png}}}
\newcommand{\orcidiconOtb}{\href{https://orcid.org/0009-0008-3903-2268}{\includegraphics[scale=0.1]{orcidID128.png}}}
\newcommand{\orcidiconOba}{\href{https://orcid.org/0000-0003-2523-3858}{\includegraphics[scale=0.1]{orcidID128.png}}}

\author{Fatih E. Bilgen\orcidiconFeb,~\IEEEmembership{Graduate Student Member,~IEEE},
        A. Sila Okcu\orcidiconAso,
        O. Tansel Baydas\orcidiconOtb,~\IEEEmembership{Graduate Student Member,~IEEE},
        and Ozgur B. Akan\orcidiconOba,~\IEEEmembership{Fellow,~IEEE}             
        \thanks{The authors are with  the Internet of Everything (IoE) Group, Electrical Engineering Division, Department of Engineering, University of Cambridge, Cambridge CB3 0FA, UK 
        (e-mails: \{feb49, aso32, otb26, oba21\}@cam.ac.uk).}
        \thanks{O. B. Akan is also with the Center for neXt-generation Communications (CXC), Department of Electrical and Electronics Engineering, Ko\c{c} University, Istanbul 34450, Turkey (e-mail: akan@ku.edu.tr).}
}

\begin{document}
\maketitle

\begin{abstract}
    Intelligent Reflecting Surfaces (IRS) are anticipated to serve as a key cornerstone of future wireless networks, providing an unmatched capability to deterministically shape electromagnetic wave propagation. Despite this potential, most existing research still considers the IRS merely as a standalone physical-layer component, controlled by transmitters. However, as networks grow to encompass a massive number of these surfaces and a massive number of transmitters wishing to use them, this transmitter-centric design encounters substantial challenges. To overcome this challenge, we propose the Internet of IRS (IoIRS), an architecture that reconceives the IRS not just as a passive reflecting surface, but as a connected, hybrid entity functioning across both the physical layer and upper network layers. We present the conceptual framework and a preliminary protocol suite necessary to integrate these surfaces into the higher network layers. We conclude by examining how IoIRS architectures could be applied in practice, as their deployment will be essential for fully realizing the capabilities of future wireless networks.
\end{abstract}


\section{Introduction}

The trajectory of wireless communication systems towards the sixth generation (6G) is characterized by an unprecedented demand for spectral efficiency, ubiquitous coverage, and energy sustainability. As the industry pivots from the sub-6 GHz bands used in 5G toward the millimeter-wave (mmWave) and terahertz (THz) spectrum, the fundamental physics of propagation changes \cite{saad_2020_vision}. These high-frequency bands offer vast contiguous bandwidths capable of supporting multi-gigabit data rates, yet they suffer from severe path loss, high atmospheric attenuation, and acute susceptibility to blockage. In this unforgiving landscape, the traditional reliance on direct Line-of-Sight (LoS) or naturally occurring scattering is no longer sufficient to design next generation communication systems. 

Intelligent Reflecting Surfaces (IRS), alternatively known as Reconfigurable Intelligent Surfaces (RIS), have emerged as a transformative solution \cite{wu_2019_intelligent}. By integrating large arrays of low-cost, passive metamaterials into the environment, IRS technology turns the wireless channel from a passive, probabilistic medium into a programmable, deterministic entity. Unlike traditional relays that actively regenerate signals or simply amplify them, the fundamental premise of IRS is the modification of the wireless channel impulse response through the intelligent manipulation of incident electromagnetic waves \cite{basar_2019_wireless}. This capability effectively introduces a controllable multipath component that can be engineered to add constructively with the direct path or create a new path entirely where none existed \cite{direnzo_2020_smart}.

While the theoretical potential of IRS to act as a smart scatterer is well established \cite{wu_2021_intelligent}, the current research trajectory largely treats it as an isolated physical layer device rather than an integrated part of a complex network. The prevailing operational model relies on a transmitter-centric design, where transmitters are responsible for knowing the exact state of the wireless environment, calculating the optimal settings for thousands of reflecting elements, and transmitting these instructions in real-time. 

\begin{figure*}[htp!]
    \centering
    \includegraphics[width=1\linewidth]{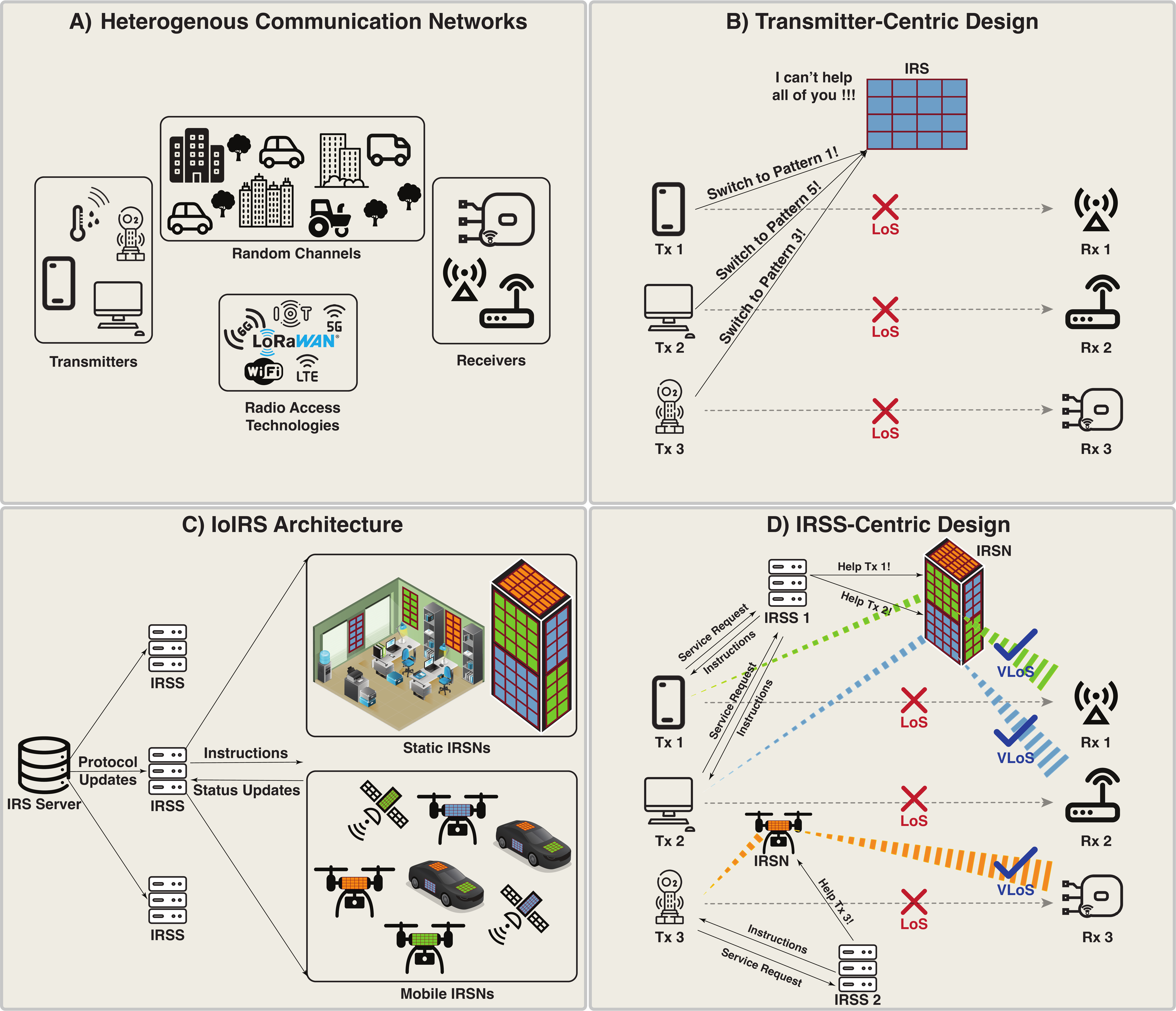}
    \caption{Overview of the proposed IoIRS framework. (a) Illustration of the complexity in future dense heterogeneous networks involving diverse transmitters and radio access technologies. (b) Operational bottlenecks in conventional Transmitter-Centric designs, where uncoordinated access attempts lead to resource contention. (c) The IoIRS hierarchical architecture, comprising an IRS Server, Stations (IRSS), and Nodes (IRSNs). IRSNs are categorized into static nodes (wall-mounted) and mobile nodes (satellite, drone, or vehicle-mounted). (d) The proposed IRSS-Centric design, where IRSS units process service requests to allocate optimal IRSNs, establishing robust Virtual Line-of-Sight (VLoS) links for blocked channels.}
    \label{fig:1}
\end{figure*}

As networks scale to accommodate massive deployments of Intelligent Reflecting Surfaces (IRS) and a dense population of transmitters contending for shared resources (Fig. \ref{fig:1}(a)), the traditional transmitter-centric design paradigm faces severe operational limitations (Fig. \ref{fig:1}(b)). This shift necessitates the design of scalable control architectures capable of effectively programming wireless channels .

To address this, we propose an architecture that reconceptualizes the IRS. Moving beyond the view of the IRS as a standalone physical layer device driven solely by the transmitter, we reimagine it as an integral network element designed to optimally program communication links. We introduce the concept of an 'Internet of IRS' (IoIRS) and present a foundational protocol suite engineered to integrate these surfaces into the broader network stack.

The remainder of this paper is organized as follows: Section \ref{section2} details the IoIRS vision and its constituent components; Section \ref{section3} explores specific use cases, demonstrating how IoIRS integration addresses prevalent issues; and finally, we conclude by outlining open challenges and potential directions for future research.

\section{Internet of IRS (IoIRS)}
\label{section2}

The Internet of IRS (IoIRS) is a transformative framework that shifts the architectural philosophy from transmitter-centric physical-layer control to a scalable distributed network-layer paradigm. In this section, we detail the three pillars of this architecture: the distributed entities that comprise the network, the operational workflow that governs their interaction, and the communication protocols that facilitate their coordination.

\subsection{Distributed Architecture and Entity Roles}

Transitioning from a centralized physical-layer topology to the IoIRS network layer requires distinct functional entities, each operating with defined scopes of autonomy. The proposed architecture distributes intelligence across four primary logical entities: the Transmitter, the IRS Server, the IRS Station (IRSS), and the IRS Node (IRSN) (Fig. \ref{fig:1}(c)).

\subsubsection{The Transmitter ($Tx$)}
The Transmitter acts as the service initiator. It is primarily in charge of generating a service request whenever it requires improvements in channel quality to achieve better performance. Moreover, It scans for and identifies the nearest accessible IRSS to act as its gateway to the IoIRS.

\subsubsection{The IRS Server}
The IRS Server functions as the global administrator. It does not intervene in real-time reflection optimization but manages the lifecycle and evolution of the network itself. To achieve this, it pushes updates to the communication structures used by all network entities, ensuring forward compatibility as the rules of the game evolve. Furthermore, it also distributes updated versions of optimization algorithms to IRSSs, enabling the network to become smarter over time without hardware replacement.

\subsubsection{The IRS Station (IRSS)}
The IRSS is the edge coordinator and decision engine. It bridges the gap between the request ($Tx$) and the physical resource (IRSNs). To maintain its state within the network, every IRSS maintains a digital identity called the IRSS ID.

The IRSS ID encapsulates the \textit{Network Parent} address of the governing IRS Server and a suite of protocol stack definitions, \textit{Protocols A, B, and C}, which establish packet structures for communication with the $Tx$, Receiver, and, IRSN respectively. Additionally, it specifies the currently loaded \textit{Optimization Engine} version for handling requests, along with the unit's physical coordinates and network identifiers, including \textit{IP} and \textit{MAC} addresses.

Operational responsibilities of an IRSS encompass \textit{Request Handling} to collect service requests from transmitters via Protocol A, followed by \textit{Receiver Localization} to identify the target location via Protocol B. The unit performs \textit{Decision Making} by executing the optimization procedure to determine the optimal subset and configuration of IRSNs for the task. Furthermore, it manages \textit{Orchestration} by sending instructions to IRSNs via Protocol C.

\subsubsection{The IRS Node (IRSN)}
The IRSN is the physical actuator, the surface itself. While it relies on the IRSS for complex calculations, it possesses sufficient local intelligence to manage its hardware state and network association. Each node includes an IRSN ID that specifies its capabilities to the IRSS it connects to.

The IRSN ID defines the surface's \textit{Physical Attributes}, including geometric dimensions, aperture configuration, and intrinsic properties, while the \textit{Power Consumption Profile} details efficiency ratings and energy metrics. It includes an \textit{Operative Frequency Index} to specify supported ranges, crucial for managing spectrally diverse clusters across fragmented bands (e.g., Sub-6 GHz, mmWave, THz) and mapping requests to appropriate hardware. The ID also lists \textit{Reflective Capabilities} covering possible patterns, a \textit{Mobility Status} flag to distinguish static versus mobile IRSNs, and \textit{Location \& Addressing} details such as physical coordinates, IP, and MAC addresses.

Operational responsibilities of an IRSN involve \textit{Execution}, where the node implements phase/amplitude patterns instructed by the IRSS with precise timing, and \textit{Status Update}, which notifies the IRSS about changes to its location or other aspects of its ID. The node performs \textit{Self-Advertisement} by broadcasting its ID to join a cluster or update its status, and engages in \textit{IRSS Solicitation} to actively search for the nearest IRSS upon deployment or displacement into a new region.

\subsection{Operational Workflow}
The operational lifecycle of an IoIRS session is a dynamic negotiation between request and resource availability. It begins when a Transmitter detects severe blockage, path loss, or multipath fading.

\subsubsection{Request and Discovery} 
The $Tx$ compiles a service packet containing current channel performance metrics and transmits a request for assistance. This request is routed to the nearest IRSS.

\subsubsection{Resource Identification}
Upon receiving the request, the IRSS queries its local registry to identify IRSNs that match the required operative frequency range. These compatible IRSNs form the \textit{Candidate Set}.

\subsubsection{Optimization and Decision Making}
The IRSS executes a multi-objective optimization procedure on the \textit{Candidate Set} to determine the optimal subset of IRSNs and their parameters. Consequently, the algorithm yields one of two distinct outcomes:
\begin{itemize}
    \item \textbf{Service Denial:} If the available IRSNs cannot improve the channel beyond the current state, or if the cost of optimization outweighs the gain, the IRSS transmits a \textit{Service Denial Packet} to the $Tx$. This action terminates the specific workflow immediately, saving signaling overhead and freeing the $Tx$ to seek alternative connection methods.
    
    \item \textbf{Service Confirmation:} The IRSS effectively determines the optimal reflection patterns and the necessary configuration for the solution. The resulting solution subset can be highly elastic, adapting to the specific constraints of the scenario. The optimization logic may deploy a single IRSN or coordinate a multi-IRSN solution. Furthermore, the architecture supports heterogeneous formations: the solution may comprise exclusively static IRSNs, exclusively mobile IRSNs, or a hybrid combination of both, depending on which configuration maximizes the objective function. This configuration dictates not only the reflection pattern modes but also the precise synchronization timestamps for switching. 
    \end{itemize}

\subsubsection{Execution (Conditional on Service Confirmation)} 
Following a confirmation of feasibility, the IRSS initiates the active control loop. It sends a confirmation packet to the $Tx$ containing timing instructions for data transmission. Simultaneously, it dispatches packets to the selected IRSNs, instructing them on which pattern to activate and the precise timing for the switch (Fig. \ref{fig:1}(d)). If the algorithm detects that physical repositioning offers superior channel gain, the solution vector includes spatial coordinates, instructing specific Mobile IRSNs to navigate to a new location. Once synchronized, the system locks in, and the $Tx$ transmits data over the newly engineered, deterministic path.

\subsection{IRS Protocol}
To ensure the IoIRS architecture integrates seamlessly into the existing ecosystem, we propose building the protocol suite directly atop the IPv6 standard. Rather than creating a non-standard encapsulation that requires specialized hardware, the IoIRS protocol utilizes the extensibility of the IPv6 header to route control information through standard network infrastructure.

\subsubsection{IPv6 Integration and Header Definition}
The IoIRS protocol packet encapsulates its specific control logic within the payload of a standard IPv6 frame. The adaptation of the \textit{IPv6 Base Header} fields is defined as follows:
\begin{itemize}
    \item \textbf{Version:} Standard IPv6.
    \item \textbf{Traffic Class:} Utilized to prioritize control signaling based on the urgency of the communication. For mission‑critical applications, routers give these requests precedence over best-effort traffic, and IRSS gives them priority over all other pending requests.
    \item \textbf{Flow Label:} Acts as a session identifier for persistent service streams. Since a single $Tx$ may manage multiple simultaneous connections with different \textit{Receivers}, the \textit{Flow Label} ensures that IRS configurations remain associated with the correct application session throughout the service duration.
    \item \textbf{Payload Length}: The size of the payload that immediately follows the \textit{IPv6 Base Header}.
    \item \textbf{Next Header:} It is a unique identifier for the IRS Protocol. This field indicates that the base header is immediately followed by an IRS packet.
    \item \textbf{Hop Limit:} A standard method applied to a packet to ensure that it does not loop endlessly within the network if it becomes lost.
    \item \textbf{Source/Destination Address:} Standard IPv6 addressing used to identify the communicating entities ($Tx$, IRSS, IRSN, $Rx$).
\end{itemize}

\subsubsection{IRS Packet Structure}
The data field following the IPv6 Base Header constitutes the IRS Packet. This is divided into a lightweight IRS Header and a variable IRS Payload.

The header contains a Message Type field. This identifier dictates the structure of the subsequent payload, supporting the interactions defined in Protocols A, B, and C.

\textit{Protocol A Payload (Tx $\leftrightarrow$ IRSS)} handles the negotiation between request and service. Its \textit{Request Payload (Tx $\to$ IRSS)} contains the Service Duration, Target Receiver IP, Transmission Parameters (e.g., Modulation Scheme, MIMO rank), and a crucial Optimization Priority Field that instructs the algorithm on weighing metrics such as SINR, Latency, Secrecy Rate, and Power Efficiency. The corresponding \textit{Response Payload (IRSS $\to$ Tx)} delivers either a Service Denial or a Service Confirmation containing precise timing synchronization data to align transmission with reflection patterns.

\textit{Protocol B Payload (IRSS $\leftrightarrow$ Rx)} acts as a feedback loop to gather target data. Through the \textit{Localization and State Payload}, the IRSS requests status updates while the Rx responds with location details used by the optimization algorithm to finalize the reflection path.

\textit{Protocol C Payload (IRSS $\leftrightarrow$ IRSN)} manages physical actuation. The \textit{Command Payload (IRSS $\to$ IRSN)} specifies reflection schedules and timestamps, uniquely instructing Mobile IRSNs to relocate via spatial coordinates. Conversely, the \textit{Status Payload (IRSN $\to$ IRSS)} provides heartbeat updates regarding availability, battery levels, and location changes to ensure the IRSS registry remains accurate.

\section{IoIRS Use Cases}
\label{section3}

The Internet of IRS (IoIRS) framework effectively decouples the control logic from the physical medium, establishing a framework that governs diverse environments ranging from dense urban canyons to orbital trajectories. It redefines environmental control by shifting from transmitter-centric design to IRSS-centric design. In this paradigm, the core challenge evolves from matching requirements with available physical resources to the efficient orchestration of distributed entities (Transmitters ($Tx$), Receivers ($Rx$), IRS Stations (IRSS), and IRS Nodes (IRSN)) across heterogeneous environments. In this section, we demonstrate how the IoIRS protocol stack adapts to two distinct domains: Next Generation Terrestrial Communications and Space Communications.

\subsection{Next Generation Terrestrial Communications}

\begin{figure*}[htp!]
    \centering
    \includegraphics[width=1\linewidth]{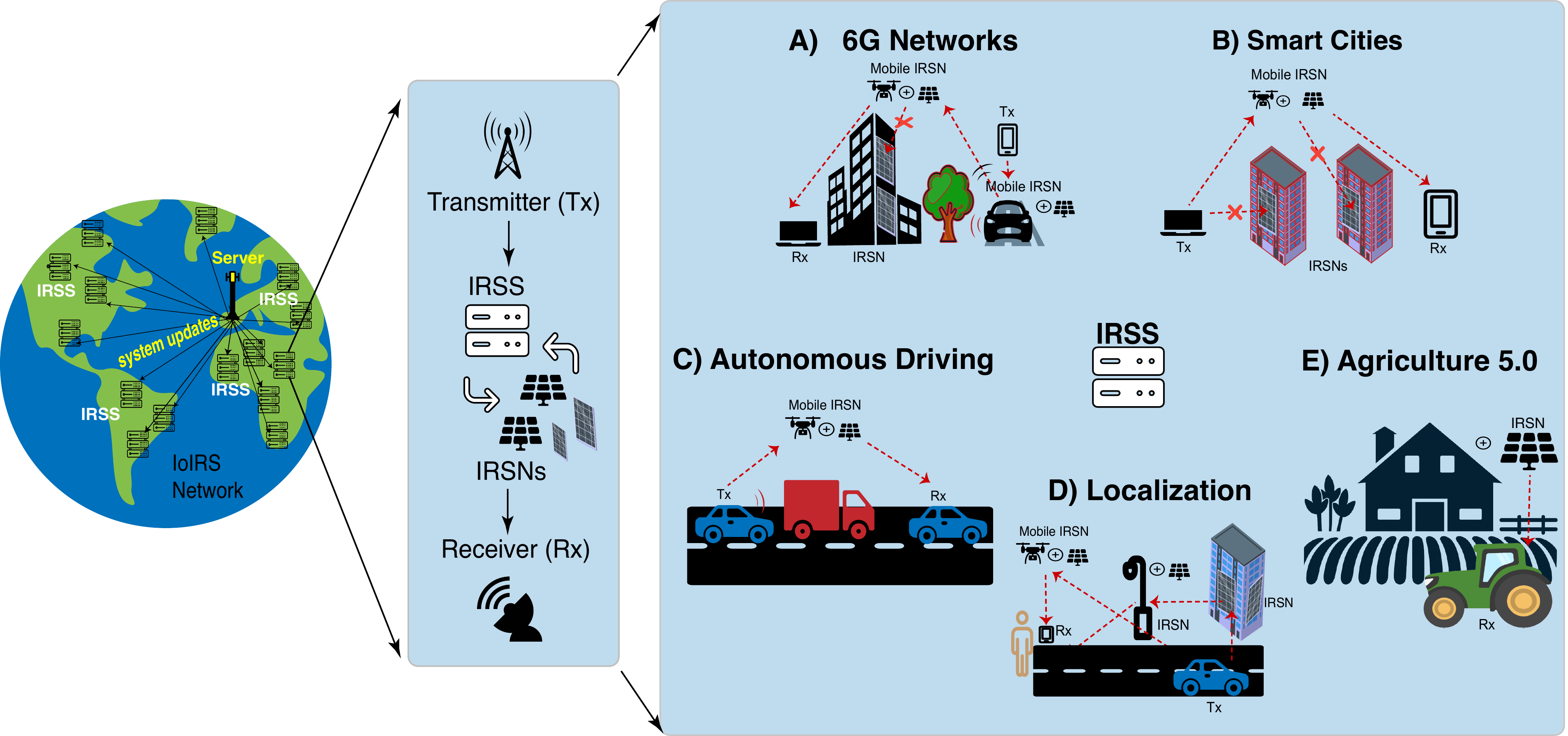}
    \caption{Illustration of the Internet of IRS (IoIRS) framework across terrestrial domains:
        \textbf{Left} shows how the IoIRS Network functions as a distributed control plane where IRS Stations (IRSS) manage global system updates and the domain specific updates are controlled through the central Server. 
        \textbf{Middle panel} visualizes key Next-Generation terrestrial scenarios supported by IoIRS by showing the operational hierarchy that separates the control logic from the physical link. A Transmitter ($Tx$) issues a request; the IRSS determines the optimal configuration and commands the IRS Nodes (IRSNs) which may be static panels or mobile units (Irs's mounted on top of mobile elements shown by \textit{addition} within right panel), to direct the channel towards the Receiver ($Rx$).
        \textbf{Right panel} visualizes the application areas:
        \textit{(A) 6G Networks}  static and drone-mounted IRSNs restore frequency links obstructed by buildings and forming virtual LoS paths under IRSS direction; 
        \textit{(B) Smart Cities} integrated metasurfaces via windows and mobile IRSNs provide outdoor-to-indoor coverage enhancement and traffic-aware routing for heterogeneous IoT flows; 
        \textbf{(C)} Autonomous Driving roadside and vehicular IRSNs maintain V2X connectivity during rapid mobility, preventing NLoS blind spots by repositioning mobile IRSNs when directed by the IRSS; 
        \textbf{(D)} Localization multiple IRSNs perform coordinated angular scans, enabling IRS-assisted cooperative positioning in urban areas where GPS signals are insufficient;
        \textbf{(E)} Agriculture 5.0 field-mounted IRSNs extend low-power rural IoT coverage by forming multi-hop reflective paths.}
    \label{fig:nextG}
\end{figure*}

The transition toward 6G envisions smart radio environments where the network adapts the physical channel rather than merely optimizing the transmission strategy \cite{wu_2021_intelligent}. To realize this vision, the infrastructure must overcome three fundamental challenges: dynamic blockage caused by rapid movement in urban environments, severe path loss at mmWave and THz frequencies, and the strict latency requirements of Ultra-Reliable Low-Latency Communication (URLLC) applications where channel coherence time fluctuates rapidly.

The IoIRS framework addresses these limitations by abstracting physical surfaces into addressable IPv6 entities. Instead of relying on static, pre-configured links, the architecture utilizes the IRSS to dynamically orchestrate IRSNs based on real-time application needs. This allows for a ``fail-fast'' or ``adapt-fast'' response to environmental volatility.

\subsubsection{6G Networks and Beyond}
In the 6G landscape, the limitation of THz signal propagation (high atmospheric attenuation) can be expected to be mitigated by the IoIRS distributed mesh. The IRSS utilizes the \textit{Protocol C} interface to coordinate multi-hop virtual LoS links, effectively daisy-chaining multiple static IRSNs to extend coverage without base station densification (Fig. \ref{fig:nextG}(a)).

Furthermore, the architecture supports Integrated Sensing and Communication (ISAC) and Wireless Power Transfer (WPT). By analyzing the Power Consumption Profile of the nodes, the IRSS can direct ambient RF energy toward power-constrained IoT devices only when the protocol payload explicitly requests WPT, thereby minimizing network-wide energy waste. For non-terrestrial integration, the IRSS treats airborne nodes (UAVs/HAPs) as standard IPv6 endpoints, seamlessly routing traffic through 3D topologies when ground-level blockage is insurmountable \cite{worka2024reconfigurable}.

\subsubsection{Smart Cities and Urban Smart Infrastructure}
The heterogeneity of smart cities poses a routing challenge: diverse traffic types (high-bandwidth video vs. massive IoT) competing for resources. The IoIRS solves this through the \textit{Traffic Class}. The IRSS acts as a traffic shaper, dynamically routing high-priority mmWave streams to metasurfaces installed on building windows (Outdoor-to-Indoor (O2I) coverage enhancement) while offloading sensor telemetry to standard sub-6GHz nodes.

Regarding O2I penetration loss, the IoIRS utilizes transparent IRSNs on smart windows. By reading the Service Request's signal requirements, the IRSS configures these surfaces to act as refractive gateways. This selective activation ensures that signals are admitted indoors only when a valid request dictates it, enhancing both coverage and security \cite{liu2022path}  (Fig. \ref{fig:nextG}(b)).

\subsubsection{Autonomous Driving}
In Vehicle-to-Everything (V2X) environments, the primary limitation is the rapid fluctuation of channel coherence time. The IoIRS addresses this via the IPv6 Flow Label. As a vehicle moves, the Flow Label ensures that the IRSS maintains session continuity, rapidly handing off control between consecutive IRSNs without re-negotiating the full service contract (Fig. \ref{fig:nextG}(c)).

To overcome Non-Line-of-Sight (NLoS) blind spots that threaten safety, the system leverages Mobile IRSNs. If static infrastructure cannot close the link, the IRSS optimization engine triggers a mobility command via Protocol C. This payload contains specific spatial coordinates, instructing a Mobile IRSN to physically navigate to a point that clears the obstruction. This capability transforms the physical environment from a static constraint into a reconfigurable variable \cite{naaz2024empowering}.

\subsubsection{IRS-Assisted Localization in NextG Settings}
In urban canyons where GPS signals are degraded, the IoIRS repurposes the communication link for positioning.

A device initiates a localization-specific Service Request. The IRSS commands multiple IRSNs to perform angular sweeps. By aggregating the feedback from the Receiver ($Rx$), specifically the Time-of-Flight (ToF) and Angle-of-Arrival (AoA) data encapsulated in the response payload, the IRSS performs cooperative triangulation. Unlike passive GPS, this method allows the network to ``hunt'' for the user, improving accuracy by recruiting additional IRSNs until the positioning error bound defined in the request is satisfied (Fig. \ref{fig:nextG}(d)).

\subsubsection{Agriculture Connectivity with IRS Assistance}
Rural agriculture faces unique challenges such as sparse base stations, long propagation distances, seasonal foliage blockage, and ultra-low power sensing constraints \cite{alsarayreh2023intelligent}.

When a soil sensor, canopy camera, or livestock tracker detects link degradation below its SINR threshold, it issues a Service Request specifying its target quality and maximum transmit power. If a single IRSN cannot satisfy these constraints, the IRSS transitions to multi-hop mode, constructing a daisy-chain reflective path through neighboring surfaces. Each intermediate IRSN adjusts its phase configuration to maintain coherent signal propagation across the chain.

This approach provides three key benefits including extended coverage without additional base stations, guaranteed uplink delivery from power-constrained devices operating within strict power budgets, and adaptive reconfiguration as seasonal foliage or weather conditions alter the channel. The framework directly supports remote sensing, environmental monitoring, and precision agriculture IoT using standard NextG protocol semantics (Fig. \ref{fig:nextG}(e)).

\subsection{Space Communications}

\begin{figure*}[hbp!]
    \centering
    \includegraphics[width=1\linewidth]{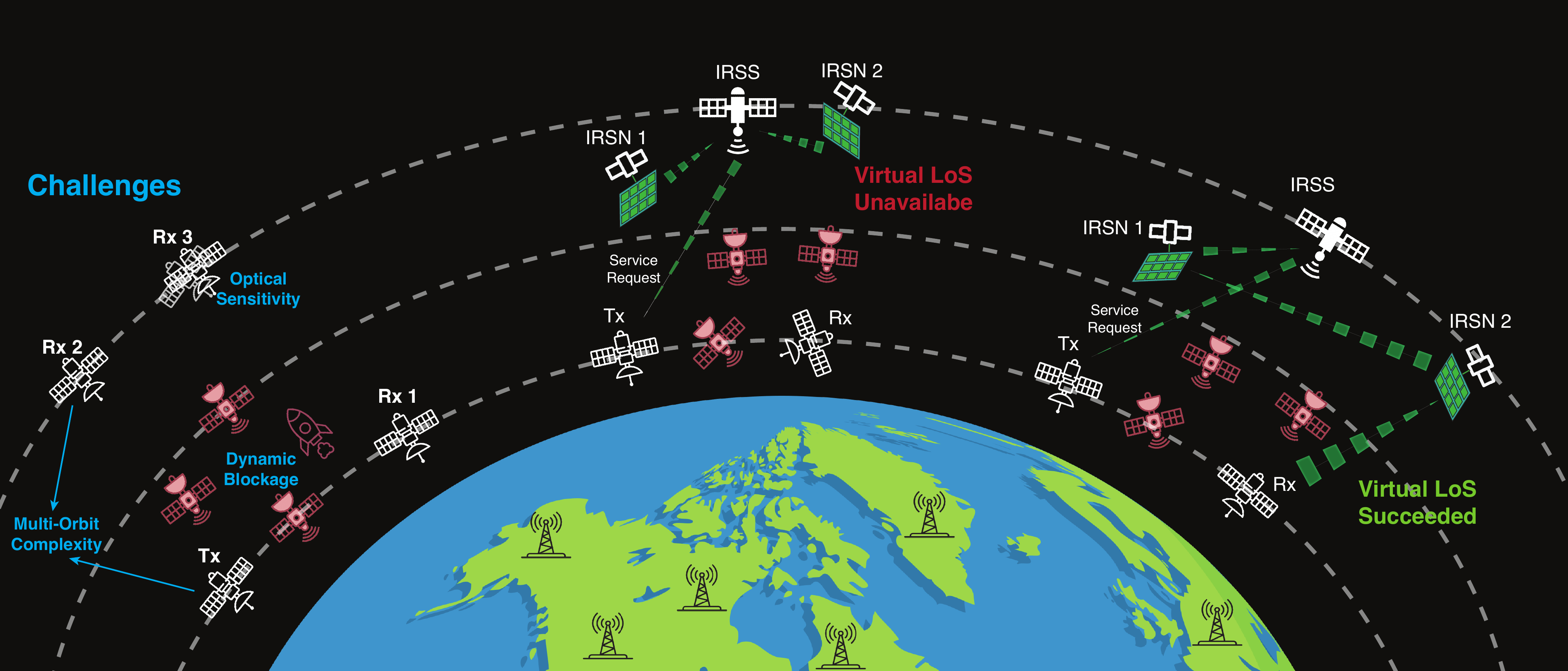}
    \caption{Space IoIRS: The left side illustrates potential challenges, while the remaining portion shows the IRS-enabled solution. When a blockage occurs, LoS link becomes unavailable, and the IRS network activates to assist the communication link. If locations of the IRSNs are not suitable, IRSNs are not feasible and they remains inactive. Once sufficient location change realized, the IRS network provides a virtual LoS path for the transmitter, enabling reliable communication with improved SINR at the receiver.}
    \label{fig:space}
\end{figure*}

The evolution of non-terrestrial networks (NTN) is shifting from simple models to advanced architectures utilizing Inter-Satellite Links (ISL). This creates an independent orbital mesh allowing global data circulation. However, maintaining this mesh requires navigating a tiered defense against severe physical and topological obstacles:
\begin{itemize}
    \item \textbf{Dynamic Obstructions:} Connections are frequently blocked by Earth eclipses or disrupted by third-satellite crossings that interrupt the active Line-of-Sight (LoS).
    \item \textbf{Multi-Orbit Complexity:} Inter-plane links between LEO, MEO, and GEO involve constant changes in relative distance and angles, inducing severe Doppler shifts that degrade synchronization.
    \item \textbf{Optical Sensitivity:} The move toward Optical Wireless Communication (OWC) demands extreme precision. Links become intolerant to pointing error variance caused by platform micro-vibrations.
\end{itemize}

To address these problems, The IoIRS framework proposes a proactive solution. In the orbital domain, the IoIRS protocol operates under constraints distinct from terrestrial systems. Here, the primary variable is not random blockage, but orbital mechanics.
\begin{itemize}
    \item \textbf{Predictive Collaboration:} Unlike terrestrial scenarios where the Candidate Set is discovered opportunistically, the Space-based IRSS utilizes orbital ephemeris data to predict the locations of IRSNs. The Service Request payload from a satellite $Tx$ triggers a ``look-ahead'' optimization. The IRSS computes handovers between IRSNs before the LoS is lost, ensuring continuous coverage across orbital planes.
    \item \textbf{Orbital Latency:} In the Service Request, latency defines the coherence time of the channel relative to the satellite's velocity. If the angular velocity of the LEO satellite exceeds the switching speed of a local IRSN, the IRSS rejects the local request.
    \item \textbf{Hardware differentiation:} Space networks increasingly hybridize RF and Optical links. The spectrum information in the IRSN ID is critical here, ensuring that a request for an optical ISL is never routed to an RF-based IRSN.
    \item \textbf{Energy Budget:} Spaceborne nodes are severely power-constrained. The energy budget in the protocol restricts the IRSS from selecting solutions that would drain a node's station-keeping batteries.
\end{itemize}

\subsubsection{Graph-Based Routing and Topology Management}
In the IoIRS architecture, the IRSS maintains a dynamic graph representation of the network to manage routing. In this abstraction, vertices represent entities ($Tx$, $Rx$, and IRSNs), while edges represent potential LoS links validated by the status updates. This graph-theoretic approach enables the IRSS to execute complex routing strategies:
\begin{itemize}
    \item \textbf{Topology Shaping}: The algorithm can virtually transform a sparse linear chain of satellites into a fully connected graph by activating specific IRSNs to bridge gaps \cite{9868343}.
    \item \textbf{Secure Routing}: When the Service Request includes a Secrecy Capacity constraint, the IRSS modifies the graph weights. It selects a path of IRSNs that maximizes the signal at the legitimate Receiver while nulling the signal in the direction of known eavesdroppers, effectively routing data through a ``secure tunnel'' in open space.
\end{itemize}

\subsubsection{Machine Learning and Navigation in IRS-Assisted Space Networks}
Space networks face a conflict between the need for complex optimization and the limited onboard processing power of satellites. The IoIRS architecture resolves this through distributed intelligence and the definitions laid out in the IRS Station entity.

As satellites shift toward edge computing, transmitting raw training data to Earth is inefficient. Instead, satellites issue a Service Request for model aggregation to utilize Over-the-Air Federated Learning (OTA-FL). The IRSS identifies a cluster of IRSNs to act as an analog computation layer. By adjusting phases, the IRSNs align the signals from multiple satellites so they sum coherently in the air. This allows the network to aggregate FL model updates instantly without processing individual streams, with the IRSN acting as a passive Doppler-compensating relay.

Also, in deep space where GPS/GNSS is unavailable, the IoIRS provides a fallback navigation solution. A satellite ($Tx$) sends a specific Localization Request. The IRSS initiates a Target Discovery Phase, commanding IRSNs to perform angular sweeps. By collecting Angle-of-Arrival (AoA) and Doppler history from the reflected signals, the IRSS triangulates the satellite's exact position and velocity.

\section{Challenges and Future Directions}

The IoIRS framework expands IRS functionality from passive reflectors to distributed, cooperative networks. This introduces new capabilities but also technical challenges. Below, we present the key challenges and corresponding future research directions across domains.

\subsection{Next-Generation Terrestrial Communications}

IoIRS operation in dense or mobile environments is constrained by several factors. First, CSI acquisition becomes expensive as large IRS panels and high mobility shorten coherence times, pilot-based cascaded estimation is impractical without geometry-assisted or learning-driven methods. Second, passive reflection imposes double path loss, and although IoIRS multi-hop routing overcomes this, path-loss compounding limits scalability unless hybrid passive–active designs are explored. Real-time beam alignment is difficult for vehicular or UAV-mounted IRS units because orientation changes and Doppler shifts invalidate static configurations. Control signaling latency remains bottleneck since optimal IRS states may expire before being applied in fast-fading channels, motivating distributed or locally adaptive IRS control.

Large-scale deployments must also confront hardware impairments, element-level bandwidth limits, and interference management; even though IoIRS negotiates collaboration, distributed clustering and scheduling remain open problems. Environmental robustness is another concern, as weather, vibration, and mechanical wear affect reflection performance, requiring calibration-free or self-calibrating surfaces. Near-field operation with ultra-large surfaces complicates modeling due to spherical wavefronts, increasing CSI estimation complexity. Finally, programmability introduces security threats such as malicious reconfiguration, eavesdropping, or poisoning, while ISAC demands multimodal hardware and sophisticated waveform design. Power consumption of IRS controllers and sensors remains critical in IoT and mobile settings.

Key opportunities include geometry-assisted CSI prediction, hybrid IRS architectures that combine passive reflection with low-power amplification, and distributed optimization algorithms that allow IRS clusters to self-organize. Predictive beam tracking and local autonomy for mobile IRS units will improve robustness. Secure IoIRS control channels, encrypted descriptors, and anomaly detection are needed for adversarial resilience. Integrated sensing–communication prototypes, energy-aware IRS protocols, and selective element activation can reduce power consumption.

\subsection{Space Communications}
Spaceborne IoIRS must satisfy size, weight, and power constraints and withstand radiation, thermal cycling, and vacuum environments while supporting rapid reconfiguration for severe Doppler compensation. High-latency control links complicate IRS coordination, requiring decentralized or partially autonomous control to maintain synchronization across rapidly moving constellation topologies. Dynamic satellite graphs demand learning models that generalize to node failures, new launches, and orbit drift without extensive retraining. Passive beam tracking is further challenged by large round-trip delays that undermine pilot-based feedback, necessitating geometry-driven or blind estimation techniques. Energy limitations constrain learning, Doppler compensation, and coordination tasks, requiring joint optimization of communication quality and onboard processing while respecting strict power budgets.

Promising future directions include radiation-hardened metasurface technologies with non-volatile phase states that reduce control signaling, distributed graph-based IRS coordination algorithms that exploit orbital predictors, and cross-layer AI frameworks that co-design resource allocation with federated learning under energy constraints. Geometry-aware IRS configuration driven by orbital mechanics rather than feedback-heavy schemes can overcome latency limitations while enabling autonomous operation during temporary ground station disconnections.

\subsection{Agriculture 5.0 and Bio-Hybrid Systems}
A promising future direction involves bio-integrated IoIRS nodes equipped with ultrasonic MEMS microphones and gas sensors to detect plant stress emissions, acoustic clicks and VOCs released during drought, pest attacks, or mechanical damage \cite{kilic_2025_information}. These hybrid units would convert biological stress cues into digital Service Requests and redistribute them through the IoIRS network, extending plants' natural communication radius from meters to entire fields. This enables early-warning systems that allow neighboring plants to pre-activate defense mechanisms without synthetic chemicals. This integration defines Agriculture 5.0, where electromagnetic connectivity, biological signal relaying, and environmental sensing operate as a unified ecosystem. Future work should explore energy-autonomous field deployments combining IRS with solar harvesting, RF energy transfer, and low-power techniques to enable operation of remote agricultural sensors\cite{alsarayreh2023intelligent}.

\section{Conclusions}
\label{section5}
The realization of ubiquitous connectivity requires more than just advanced physical materials; it demands a fundamental rethinking of how network elements interact. While Intelligent Reflecting Surfaces (IRS) offer the physical capability to reshape wireless channels, adhering to a transmitter-centric design stifles their potential in dense, large-scale deployments. In this article, we have proposed the Internet of IRS (IoIRS) as a necessary evolution, transforming IRS from passive endpoints into active, hybrid network nodes. By defining a preliminary protocol suite and integrating these structures into the upper layers of the network stack, we can alleviate the processing burden on transmitters and enable truly scalable, autonomous channel control. As we move forward, the research focus must shift from isolated physical-layer optimization to the standardization of these network protocols, ensuring that the programmable wireless environment becomes a practical reality rather than just a theoretical promise.

\bibliographystyle{ieeetr}
\bibliography{references.bib}

\end{document}